\documentclass[11pt,twoside]{article}
\usepackage{newpasp}

\input{psfig}
\newcommand{\HI}{H{\,\small I}}

\index{Morganti, R.}
\index{Oosterloo, T.A.}
\index{Tadhunter, C.N.}
\index{Wills, K.A.}

\begin{document}

\title{Neutral and ionized gas distribution in and around the radio galaxy Coma A}
\author{R. Morganti, T. Oosterloo}
\affil{ASTRON - NFRA, Postbus 2, 7990 AA
Dwingeloo, The Netherlands}
\author{C.N. Tadhunter, K. A. Wills}
\affil{Dep. Physics and Astronomy,
University of Sheffield,  S7 3RH, UK}


\begin{abstract}

\HI\ absorption has been detected with the WSRT against {\sl both lobes} of
the radio galaxy Coma~A.  This radio galaxy could be expanding in a
particularly gas rich environment, perhaps the result from interactions/mergers
between the dominant giant galaxy (associated with the radio galaxy) and less
massive galaxies in the same group. 
\end{abstract}




Powerful radio galaxies are frequently associated with kpc-sized emission line
nebulosities, extending up to tens of kpc from the nucleus.  These
nebulosities can be either lit up by interactions between the radio jet/lobe
and the environment or photoionized by anisotropic UV radiation from the
active nucleus. Little is known in radio galaxies to what extent this ionized
gas is related to the overall ISM.  A key element for studies of these
extended emission line regions is to know the intrinsic distribution of the
gas in and around the galaxy, i.e.  to know the distribution of the neutral
gas together with that of the ionized gas.  Coma~A is an ideal object for this
study.  It is a well-known radio galaxy (\( z=0.0857 \)) studied in detail by
van Breugel et al. (1985).  The radio structure is mainly formed
by two lobes (see Fig.~1) and the total size is $\sim 65$ kpc (for \( H_{\circ
}=75 \) km s\( ^{-1} \) Mpc\( ^{-1} \), 1.45 kpc/arcsec).  A spectacular
system of interlocking emission line arcs and filaments has been observed in
H$\alpha$ (see Fig.~1) using the Taurus Tunable Filter (on the WHT, Tadhunter
et al. 2000).  A striking match between the ionized gas and the radio
structures (see Fig.~1) makes the origin of the ionized gas even more puzzling
and {\sl strongly suggests the presence of a complex interaction between the radio
structure and the gas around Coma~A}.

\section{HI observations with the Westerbork Synthesis Radio Telescope}

We have observed Coma A at the frequency of the redshifted \HI\ line (1343
MHz) with the WSRT using 10 MHz bandwidth and a velocity resolution of $\sim
20$ km s$^{-1}$.  At the resolution of our observations (about 13 arcsec), the
radio structure of Coma A appears just resolved in two structures
corresponding to the northern and southern lobes.  The preliminary results
show that we have detected \HI\ in absorption against {\sl both radio lobes}.
In both lobes the peak of the absorption is at a very low optical depth
($\sim 0.4$ \%).  This low optical depth is likely to be due (at least in part) to
dilution.  The velocity of the absorption is about 200 km s\( ^{-1} \)
redshifted for the northern lobe and between 100 and 150 km s\( ^{-1} \)
blueshifted for the southern lobe compared to the systemic velocity.  The FWHM
is between 150 and 200 km s\( ^{-1} \).  Assuming a spin temperature of 100 K,
this gives a column density of \( 5\cdot 10^{19} \) atoms cm\( ^{-2} \). 
Velocities of the ionized gas derived from a slit aligned with the radio axis
show the same trend with position.  Thus, {\sl the HI appears kinematically
associated with the ionized gas}. 

\centerline{\psfig{figure=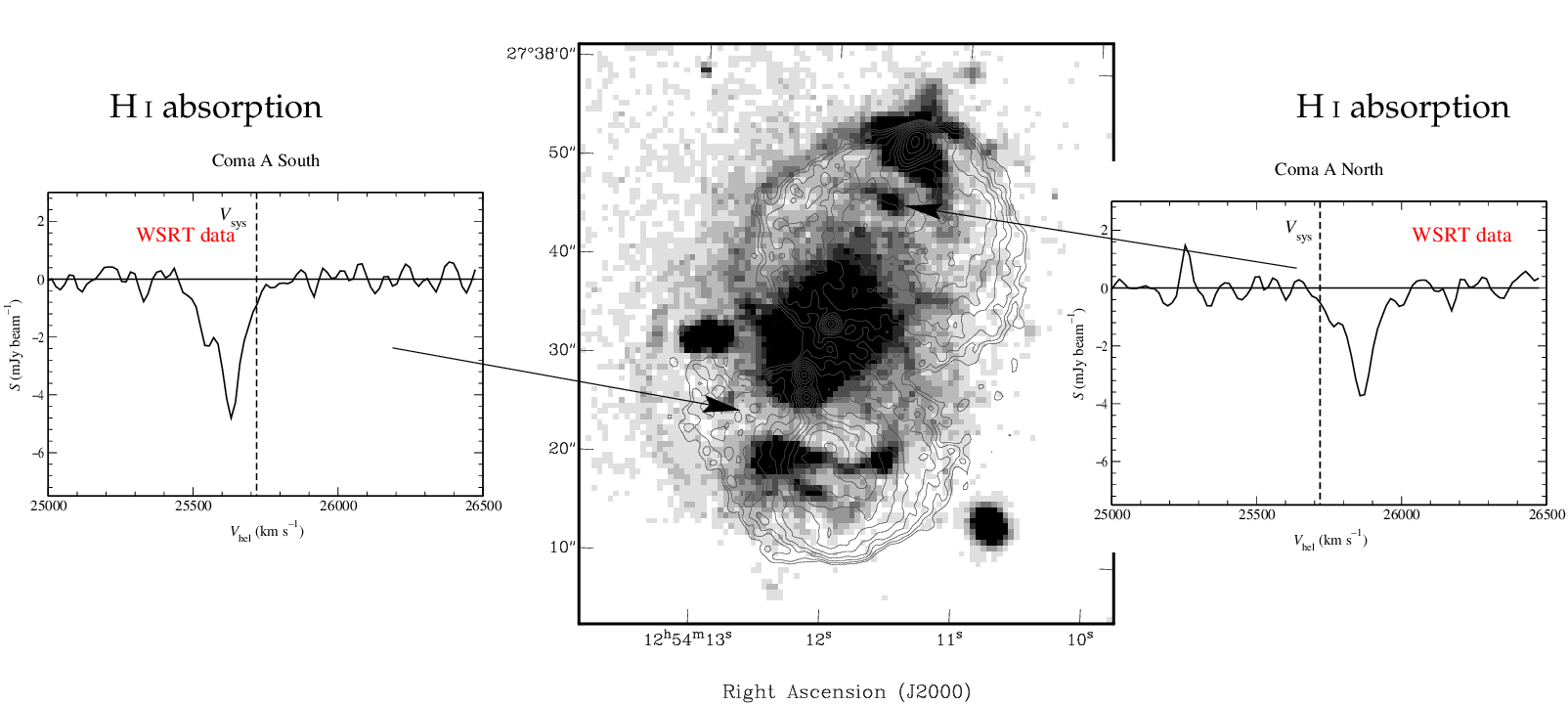,height=6.0cm,angle=-90}}
{\it {\bf Fig.1} Panel showing the approximate location of the \HI\ absorption.
In the plots, the dashed line  represents the systemic velocity.
In the middle, is the H$\alpha$ image with superimposed the contours from a
20cm VLA image.}

\subsection{Preliminary conclusions}

One of the very few objects in which spatially extended 21-cm \HI\ absorption
has been detected against the radio lobes is the central radio galaxy in the
cooling flow cluster Abell 2597 (O'Dea et al. 1994).  This has been
interpreted as indication that the radio lobes are surrounded by a collection
of clouds containing both neutral and ionized gas.
The same could apply to Coma A, with the main difference that {\sl the scale of the
phenomena is larger} (several tens of kpc instead of few kpc).  Indeed, in the
case of Coma A, the complicated H\( \alpha \) structure has been explained by
a particularly gas rich environment surrounding the galaxy and that the close
morphological association is due to the radio lobes expanding into this
gas-rich environment (Tadhunter et al. 2000). The detected \HI\ would just
be part of this system.

Coma A could be the result of interactions/mergers between the dominant giant
galaxy and less massive galaxies in the same group.  Although we cannot yet
say how spatially extended the observed absorption is, it is quite likely
that Coma A indeed possesses extended \HI\ structures.



\end{document}